\documentstyle[preprint,aps]{revtex}

\begin{document}

\draft

\title{Renormalization of the charged scalar field in curved space}

\author{Rhett Herman\cite{her} and William A. Hiscock\cite{his}}
\address{Department of Physics, Montana State University, Bozeman, Montana
59717}

\date{September 6, 1995}

\maketitle
\begin{abstract}
The DeWitt-Schwinger proper time point-splitting procedure is applied to
a massive complex scalar field with arbitrary curvature coupling
interacting with a classical electromagnetic field in a general curved
spacetime. The scalar field current is found to have a linear divergence.
The presence of the external background gauge field is found to modify the
stress-energy tensor results of Christensen for the neutral scalar field by
adding terms of the form $(eF)^2$ to the logarithmic counterterms.
These results are shown to be expected from an analysis of the degree of
divergence of scalar quantum electrodynamics.
\end{abstract}


\section{Introduction}

In the study of quantized fields in curved spacetimes, the use of geodesic
point-splitting as a regularization method for quantities such as the vacuum
polarization $\langle \phi^2 \rangle$ and the stress-energy tensor
$\langle T_{\mu \nu} \rangle$ has been shown to be a robust and trustworthy
method \cite{BD}. The detailed method of point-splitting regularization was
developed from the DeWitt-Schwinger proper time method of calculating the
Feynman Green function by Christensen \cite{Christensen,Christensenthesis}
for the stress-energy
tensor of a massive real scalar field. In this paper, we extend
Christensen's work to develop the point-splitting counterterms necessary
to regularize the current four-vector and stress-energy tensor for a
massive complex charged scalar field with arbitrary curvature coupling
in a general curved spacetime background with a nonzero background
classical electromagnetic field. These results will allow, for the first time,
calculation of both the expectation value of the current,
$\langle j^\mu \rangle$, and the stress-energy tensor, $\langle T_{\mu \nu}
\rangle$, of a charged quantized field in a fixed background spacetime
containing an electromagnetic field. In addition, our results may also be
used to consider the combined Einstein-Maxwell semiclassical
backreaction problem, where the gravitational and electromagnetic fields
are treated classically, and their sources are taken to be quantized fields:
\begin{equation}
 	G_{\mu\nu}=8\pi\langle T_{\mu\nu}\rangle,
\label{Eq1} \end{equation}
and,
\begin{equation}
	F^{\mu\nu}{}_{;\nu}=4 \pi \langle j^\mu \rangle.
\label{Eq2} \end{equation}

Our primary motivation for this work is to apply these results in the
study of charged quantized fields in the spacetime of a charged spherical
black hole, which would be described classically by the Reissner-Nordstr\"{o}m
metric. The electromagnetic field of the Reissner-Nordstr\"{o}m black hole
gives rise to such exotic structures in the black hole interior as Cauchy
horizons, timelike singularities, and an apparent tunnel to other
asymptotically flat spacetimes.

The key physical question is whether the analytically extended
Reissner-Nordstr\"{o}m solution correctly describes the interior of a charged
spherical black hole. The Cauchy horizon has been shown to be both
classically \cite{pn,sp,mcn,matz,gur,chan} and quantum mechanically unstable
\cite{hi,ndb,hi2}. The metric backreaction to the classical instability has
been
shown to result in the Cauchy horizon becoming a curvature singularity, via
the so-called "mass inflation" process \cite{Hiscock1,Poisson}. It is, however,
somewhat unclear whether this singularity is sufficiently strong to act as
a true ``edge'' to the spacetime, enforcing strong cosmic censorship
\cite{Ori,HH1}.

The studies of Cauchy horizon instability and mass inflation deal with
the evolution of uncharged fields on an initially Reissner-Nordstr\"{o}m
background. Yet, since it is the existence of the electromagnetic field
which causes the exotic internal structures such as the Cauchy horizons to
exist, one should surely study the stability and evolution of the
electromagnetic field in these solutions. In particular, since extremely
strong electric or magnetic fields will be encountered in the black hole
interior, it is appropriate to study the pair creation and vacuum polarization
of charged quantized fields in these spacetimes. Gibbons \cite{Gibbons} has
studied the effects of thermal particle production on charged black holes,
while Davies \cite{Davies} has considered the thermodynamic implications of how
the charge of the black hole affects its evaporation. However, both of these
studies were performed in the context of a fixed background (constant
mass, $M$, and charge, $Q$) spacetime.

Some studies have been performed wherein the electromagnetic field of the
black hole is allowed to create charged pairs whose field serves to
modify the electromagnetic field that created them. Hiscock and Weems
\cite{HW1}
considered the loss of charge and mass for the exteriors of
charged black holes of large initial mass. They assumed adiabatic evolution,
so that the spacetime could be described by a sequence of
Reissner-Nordstr\"{o}m
metrics with the mass and charge of the black hole being slowly varying
functions of time.  Integrating these equations, they investigated the
dynamical future of the discharging, evaporating black hole, but their
treatment
was limited to the region exterior to the event horizon.

Several studies have attempted to model the evolution of the interior
electromagnetic field and metric.
Novikov and Starobinski\'{i} \cite{NS} have studied the dynamical
evolution of the electric field within the outer horizon. In their model,
the electric field serves as the source of its own demise through the
production of electron-positron pairs. The effect of these pairs on the
interior spacetime is such that it seems to evolve from an initial
Reissner-Nordstr\"{o}m state to a final state which is uncharged and (in
broad terms) Schwarzschild-like. Their model assumes an initial
Reissner-Nordstr\"{o}m geometry on a spacelike slice inside the outer
event horizon with the pair production being allowed to modify the
electric field to the future of that initial surface.

So far these studies have been done using many simplifying
assumptions. These assumptions, while giving intuitively satisfying
results, are still approximations to the true dynamical
equations. They have often used the Schwinger formula
\cite{Schwinger} for the number of charged pairs created per unit
four-volume of spacetime as the source of their pair creation. This
formula was derived with its own simplifying assumptions of a static and
uniform electric field, and a flat background spacetime. And while it is
considered valid in the regime of small curvature and slowly varying
electromagnetic fields, it is still, in essence, a flat-space result.

Clearly, a more complete treatment of the problem would involve choosing an
initial spacelike surface consistent with an asymptotic
Reissner-Nordstr\"{o}m spacetime. Dynamical equations would then be
allowed to evolve these initial conditions into the future, determining
the true geometry through the use of both the semi-classical Einstein
field equations and the Maxwell equations, Eqs.(1-2). In this context
the problem of the Cauchy horizon would change from a question of whether
it is unstable to some perturbation into a question of whether it is formed
at all in a self-consistent treatment including semiclassical effects.

Recently, Anderson, Hiscock, and Samuel \cite{AHS} have described a
method for numerically computing the vacuum expectation value
of the stress tensor for
quantized scalar fields in static spherically symmetric spacetimes with
arbitrary mass and curvature coupling. Their method is fully renormalized
in that they combine the DeWitt-Schwinger counterterms for the stress
tensor of the real scalar field derived by Christensen \cite{Christensen}
with the expressions for the components of the unrenormalized stress
tensor. With this renormalization, their computational scheme may be
carried to arbitrary numerical precision. This method is completely
general and may be extended to the calculation of the expectation value
of the current due to creation of charged pairs by the electromagnetic field
of the charged black hole.

To begin the study of the semiclassical dynamics of a charged spherical
black hole interior, in this paper we develop the theory of point-splitting
regularization of the four-current vector and stress-energy tensor of a
charged massive complex scalar field with arbitrary curvature coupling.
The background spacetime is arbitrary, as is the background classical
electromagnetic field. The complex scalar field is chosen for study in order
to better make contact with existing point-splitting results, which have
largely been derived only for scalar fields, and to avoid (for the present)
the
added complications of having to deal with spin.

In Sec. \ref{sec:semiclassical}, the equations for the expectation values
of the current and stress-energy tensor equations are presented in the form
required for the point-splitting procedure, revealing the Hadamard elementary
function, $G^{(1)}(x,x')$, and its gauge covariant derivatives as the
key quantities to analyze.
In Sec. \ref{sec:green}, the ansatz of DeWitt used
by Christensen is used to yield recursion relations of the familiar form. The
presence of the gauge field in this case is seen to modify the
original recursion
relations such that they assume a gauge invariant form.
Sec. \ref{sec:coincidence} contains a listing of the necessary
coincidence limits required in the point-splitting expansions of the
bitensors involved in the vacuum expectation value (VEV) of the current
and stress tensor.
Sec. \ref{sec:expansions} then uses these coincidence limits to expand
the various bitensors appearing in $G^{(1)}(x,x')$ and its derivatives in
power series about the fixed point $x$. The series involve powers of the
separation vector connecting the point $x$ to the nearby point $x'$.
In Sec. \ref{sec:results}, the quantities derived in the preceding
sections are combined to construct the divergences of the VEV of the
current and the stress tensor. The complex scalar field current is seen
to have a linear divergence, while the presence of the electromagnetic field
adds solely to the logarithmic divergence of the stress tensor. Sec.
\ref{sec:discussion} briefly discusses why these results are consistent
with those expected from an analysis of scalar quantum electrodynamics.

\section{Semi-classical point-separated current and stress tensor}
\label{sec:semiclassical}

A complex scalar field coupled to the electromagnetic field in an arbitrary
curved background is described by the action functional \cite{MTW},
\begin{equation}
 S[\phi,A_\mu] = -{1\over 2}\int { g^{1 \over 2}
   \left[ (D_\mu \phi)(D^\mu \phi)^* + (m^2 + \xi R)\phi\phi^* -
   {1\over 4}F_{\mu\nu}F^{\mu\nu} \right] d^4 x },
\label{action} \end{equation}
where $\phi(x)={1\over \sqrt{2}}\left(\phi_{1}(x) +
i\phi_{2}(x)\right)$ is the complex scalar field, $g$ is the negative of the
determinant of the metric $g_{\mu\nu}$,
$D_{\mu} \equiv (\nabla_\mu - ieA_\mu)$ is the gauge
covariant derivative, $A^\mu$ is the classical electromagnetic
vector potential, $e$ is the coupling between the complex scalar and the
electromagnetic fields, $m$  is the mass of the complex scalar field,
$\xi$ is the scalar curvature coupling, and $R$ is the scalar curvature.
The wave equation obtained by varying the action in Eq.(\ref{action}) with
respect to $\phi^*$ is
\begin{equation}
  \phi_{|\mu}{}^{\mu}-(m^2+\xi R)\phi = 0 ,  \label{wave}
\end{equation}
where the vertical slash denotes gauge covariant differentiation.
The classical current equation obtained from Eq.(\ref{action}) is
\begin{equation}
  j^\mu \equiv {1 \over {4\pi}}F^{\mu\nu}{}_{;\nu} = {1\over 2}ie\left[
  \{D^{\mu}\phi,\phi^*\}-\{D^{\mu}\phi,\phi^*\}^*\right],
\label{classcurrent} \end{equation}
where the braces \{\} denote the anticommutator. The components of the
classical stress-energy tensor are given by,
\begin{eqnarray}
  T^{\mu\nu} & = & {1\over 2} \lbrack
    {1\over 2}(1-2\xi)\{\phi^{|\mu},\phi^{*|\nu}\}+
   {1\over 2}(2\xi-{1\over 2})g^{\mu\nu}\{\phi_{|\sigma},\phi^{|\sigma}\}
   \nonumber \\
  && {}-\xi\{\phi^{|\mu\nu},\phi^{*}\}+
       \xi g^{\mu\nu}\{\phi_{|\sigma}{}^{\sigma},\phi^* \} \nonumber\\
  && +{1\over 2}\xi(R^{\mu\nu}-{1\over 2}g^{\mu\nu}R)\{\phi,\phi^*\}-
       {1\over 4}m^2 g^{\mu\nu}\{\phi,\phi^*\} + c.c.\rbrack,
\label{classstress}
\end{eqnarray}
where $c.c.$ denotes the complex conjugate of all of the previous terms.
The anti-commutators above arise from symmetrizing with respect to the
fields $\phi$ and $\phi^*$. As usual, the transition from
classical to quantum fields is made by replacing each classical field $\phi(x)$
with a field operator $\underline{\phi}(x)$. Then, following Christensen,
the first field operator $\underline\phi(x)$ in each bracket in
Eq.(\ref{classcurrent}) is moved to the nearby
point $x'$ and evaluated between the vacuum states $\langle out,vac|$
and $|in,vac\rangle$. A similar expression is then obtained by taking the
quantum version of Eq.(5) and moving the second field operator to the point
$x'$. The two results are then averaged.

The Hadamard function $G^{(1)}$ and its first few gauge covariant derivatives
may be written as
\begin{mathletters}
\label{allequations}
 \begin{equation}
  G^{(1)}(x,x')  \equiv
  {{\langle out,vac|\{\underline{\phi}(x),\underline{\phi}^*(x')\}
	|in,vac \rangle} \over {\langle out,vac|in,vac \rangle}},
 \label{G1}
 \end{equation}
 \begin{equation}
  G^{(1)|\mu} \equiv
  {{\langle out,vac|\{\underline{\phi}^{|\mu}(x),\underline{\phi}^*(x')\}
	|in,vac \rangle} \over {\langle out,vac|in,vac \rangle}},
 \label{G1d}
 \end{equation}
 \begin{equation}
  G^{(1)|\mu'} \equiv
  {{\langle out,vac|\{\underline{\phi}(x),\underline{\phi}^{*|\mu'}(x')\}
	|in,vac \rangle} \over {\langle out,vac|in,vac \rangle}},
 \label{G1p}
 \end{equation}
 \begin{equation}
  G^{(1)|\mu\nu} \equiv
  {{\langle out,vac|\{\underline{\phi}^{|\mu\nu}(x),\underline{\phi}^*(x')\}
	|in,vac \rangle} \over {\langle out,vac|in,vac \rangle}},
 \label{G1dd}
 \end{equation}
 \begin{equation}
  G^{(1)|\mu\nu'} \equiv
  {{\langle
out,vac|\{\underline{\phi}^{|\mu}(x),\underline{\phi}^{*|\nu'}(x')\}
	|in,vac \rangle} \over {\langle out,vac|in,vac \rangle}},
 \label{G1dp}
 \end{equation}
and
  \begin{equation}
  G^{(1)|\mu'\nu'} \equiv
  {{\langle out,vac|\{\underline{\phi}(x),\underline{\phi}^{*|\mu'\nu'}(x')\}
	|in,vac \rangle} \over {\langle out,vac|in,vac \rangle}}.
 \label{G1pp}
 \end{equation}
\end{mathletters}
With these definitions, the expectation value of the scalar field current
may be written in terms of the Hadamard function as
\begin{equation}
\langle\underline{j}^{\mu}(x)\rangle = \lim_{x'\to x}{ie\over 4}\left[
  \left( G^{(1)|\mu}+g^{\mu}{}_{\tau'}G^{(1)|\tau'} \right) -
  \left( G^{(1)|\mu}+g^{\mu}{}_{\tau'}G^{(1)|\tau'} \right)^*\right].
\label{divcurrent} \end{equation}
Note the current is guaranteed to be real. The expectation value of the
stress-energy tensor may similarly be written as
\begin{eqnarray}
 \langle\underline{T}^{\mu\nu}(x)\rangle & = & \lim_{x'\to x}Re \lbrack
   {1\over 2}({1\over 2}-\xi)
   (g^\mu{}_{\tau'}G^{(1)|\tau'\nu}+g^\nu{}_{\rho'}G^{(1)|\mu\rho'})
   +(\xi-{1\over 4})
   g^{\mu\nu}g^{\alpha\rho'}G^{(1)}{}_{|\alpha\rho'}
   \nonumber \\
 && -{1\over 2}\xi
    (G^{(1)|\mu\nu}+g^\mu{}_{\tau'}g^\nu{}_{\rho'}G^{(1)|\tau'\rho'})
  +{1\over 4}\xi g^{\mu\nu}
    (G^{(1)}{}_{|\alpha}{}^{\alpha}+
     g^\alpha{}_{\tau'}g_{\alpha\rho'}G^{(1)|\tau'\rho'})
      \nonumber                             \\
 && +{1\over 2}\xi(R^{\mu\nu}-{1\over 2}g^{\mu\nu}R)G^{(1)}-
     {1\over 4}m^2 g^{\mu\nu}G^{(1)}  \rbrack.
\label{divstress}
\end{eqnarray}

Eqs.(\ref{divcurrent}) and (\ref{divstress}) are divergent;
the following section will outline the point-splitting procedure for
isolating the infinities of $G^{(1)}$ and its derivatives needed to regularize
these expressions.

\section{Green Functions and the Recursion Relations}
\label{sec:green}
To isolate the infinities of the Hadamard elementary function
$G^{(1)}(x,x')$ as the points are brought together, we use the relation,
\begin{equation}
	G_{F}(x,x') = \overline{G}(x,x') - \frac{1}{2}iG^{(1)}(x,x'),
\end{equation}
where $G_{F}(x,x')$ is the Feynman Green function, and $\overline{G}(x,x')$
is one-half the sum of the advanced and retarded Green functions. In
coordinate space, the Feynman Green function satisfies,
\begin{equation}
	F(x) G_{F}(x,x') = -\delta(x-x') ,
\end{equation}
with the operator $F(x)$ given by $F(x)=g^\frac{1}{2}
\left[D_{\mu}D^{\mu}-(m^2+\xi R)\right]$; derivatives
are with respect to $x$. After rewriting this as the matrix equation,
\begin{equation}
  \int \langle x|\underline{F}|x''\rangle\langle
  x''|\underline{G}|x'\rangle d^{4}x'' = \langle x|-\underline{1}|x\rangle,
\end{equation}
$\underline{F}$ and $\underline{G}$ are now matrix operators.
Inserting two factors of $\underline{1}={\underline{g}}^{-\frac{1}{4}}
{\underline{g}}^\frac{1}{4}$ maintains the transformation properties of
the matrix operators and allows the matrix equation to be written as,
\begin{equation}
 {\underline g}^{1\over 4}\underline{G} {\underline{g}}^{1\over 4} =
 {{\underline{1}}\over
  -({{\underline{g}}^{{-1\over 4}}\underline{F}{\underline{g}}^{-{1\over 4}}
   +i\epsilon \underline{1} }) } = i\int_0^\infty e^{-i\underline{H}(s-0)}ds,
\label{gfmatrix} \end{equation}
with the matrix operator $\underline{H}\equiv -({{\underline{g}}^{-{1\over
4}} \underline{F} {\underline{g}}^{-\frac{1}{4}}+i\epsilon \underline{1}})$.
Taking matrix elements of Eq.(\ref{gfmatrix}) and rearranging, yields,
\begin{eqnarray}
  G(x,x') & = & i\int_0^\infty g^{-{1\over 4}}(x)
  \langle x|e^{-i\underline{H}(s-0)}|x' \rangle  g^{-\frac{1}{4}}(x') ds
  \nonumber \\
  & \equiv & i\int_0^\infty g^{-\frac{1}{4}}(x)
	\langle x,s|x',0 \rangle  g^{-\frac{1}{4}}(x') ds.
\label{DS} \end{eqnarray}
The matrix element $\langle x,s|x',0 \rangle$ obeys the
Schr\"{o}dinger-like equation,
\begin{eqnarray}
  i{\partial\over{\partial s}} \langle x,s|x',0 \rangle & = &
  \langle x,s|\underline{H}|x',0 \rangle \nonumber \\
  & = &{g^\frac{1}{4}(x)g^{-\frac{1}{4}}(x')
       \left[(m^2+\xi R) - D_{\mu}D^{\mu}\right]} \langle x,s|x',0 \rangle,
\label{schrodinger} \end{eqnarray}
with the boundary condition,
\begin{equation}
 \langle x,0|x',0\rangle=\langle x|x'\rangle=\delta(x-x'),
\label{boundary}  \end{equation}
where the infinitesimal factor $+i\epsilon$ has been dropped for brevity.
We use the same ansatz as DeWitt \cite{DeWitt} for the solution of
Eq.(\ref{schrodinger}),
\begin{equation}
  \langle x,s|x',0 \rangle = -{i \over (4\pi)^2}
	{D^{1\over 2}(x,x')\over s^2}
	exp\left[i{\sigma(x,x')\over {2s}} - i m^2 s\right] \Omega (x,x'),
\label{ansatz} \end{equation}
where the Van Vleck-Morette determinant is defined by $D(x,x')\equiv
-det(-\sigma_{;\mu\nu'})$, and
\begin{equation}
  \sigma(x,x')\equiv{1\over 2}\sigma^{|\mu}\sigma_{|\mu}
  				   ={1\over 2}\sigma^{,\mu}\sigma_{,\mu}
\label{sigma} \end{equation}
is one-half of the square of the geodesic distance between $x$ and $x'$.
The vector $\sigma^{,\mu}$ is tangent to the geodesic at the point $x$,
has length equal to the geodesic distance between the points $x$ and $x'$,
and points in the direction $x'\rightarrow x$.
We also use the identity \cite{DeWitt},
\begin{equation}
  D^{-1} \left( D\sigma^{|\mu} \right)_{|\mu} = 4.
\label{VVM} \end{equation}
It should be noted that the scalars $\sigma$ and $D$ are geometric
quantities unaffected by the presence of the gauge field. Thus their
covariant derivatives will require the Christoffel connection but will
not need any connection to the gauge field through $A^\mu$.

Substituting Eq.(\ref{ansatz}) into Eq.(\ref{schrodinger}), and using
Eq.(\ref{VVM}), a differential equation for $\Omega(x,x')$ is obtained,
\begin{equation}
  i{{\partial\Omega}\over{\partial s}}+
  {i\over s}\Omega_{|\mu}\sigma^{|\mu}=-D^{-{1\over 2}}
  (D^\frac{1}{2} \Omega)_{|\mu}{}^{\mu}+\xi R \Omega.
\label{omega} \end{equation}

We now assume that $\Omega$ may be represented by the power series,
\begin{equation}
  \Omega(x,x') = \sum_0^\infty a_{n}(x,x') (is)^n .
\label{omegaseries} \end{equation}
This may be done, so long as the gravitational and electromagnetic fields,
upon which the coefficients $a_{n}(x,x')$ depend, are slowly varying over
the infinitesimal distance between the points $x$ and $x'$. Substituting
Eq.(\ref{omegaseries}) into Eq.(\ref{omega}), and defining
$\Delta(x,x')\equiv g^{-{1\over 2}}(x) D(x,x')
g^{-{1\over 2}}(x')$, yields the recursion relations which will be used to
determine the $a_n$,
\begin{equation}
  \sigma^{|\mu} a_{0|\mu} = 0,
\label{a0recursion} \end{equation}
and
\begin{equation}
  \sigma^{|\mu} a_{n+1|\mu}+(n+1)a_{n+1} =\Delta^{-{1\over 2}}
  (\Delta^{1\over 2} a_n)_{|\mu}{}^{\mu}- \xi R a_n.
\label{anrecursion} \end{equation}
These relations are of the same form as those derived by Christensen.
The presence of the gauge field in this case now requires that all
derivatives be gauge covariant. As mentioned previously, objects such as
$\sigma^{|\mu},\Delta^{{1\over 2}|\mu\nu},R^{|\mu\nu\rho},$ etc., will
only require the Christoffel connections. Only the coefficients
$a_n(x,x'),$ which carry information about the gauge field, will require
the gauge connection $A^\mu$ in their derivatives.

Finally, after substituting Eq.(\ref{omegaseries}) and Eq.(\ref{ansatz}) into
Eq.(\ref{DS}), and performing a few straightforward but long mathematical
manipulations which will not be detailed here (see Refs.
\cite{Christensen,DeWitt} for details), we arrive at the final form for
$G^{(1)}(x,x')$, namely
\begin{eqnarray}
G^{(1)}(x,x')  =  {{\Delta^\frac{1}{2}}\over{4 \pi^2}} &  &\Biggl\{
  a_0\left[{1\over\sigma}+m^2(\gamma+\frac{1}{2}\ln|\frac{1}{2}m^2\sigma|)
		(1+\frac{1}{4}m^2\sigma + \cdots) -
		\frac{1}{2}m^2-\frac{5}{16}m^2\sigma + \cdots\right]
		\nonumber \\
&  &  -a_1\left[(\gamma+\frac{1}{2}\ln|\frac{1}{2}m^2\sigma|)
		(1+\frac{1}{2}m^2\sigma + \cdots) -
		\frac{1}{2}m^2\sigma-\cdots \right] \nonumber  \\
&  &  +a_2\sigma\left[(\gamma+\frac{1}{2}\ln|\frac{1}{2}m^2\sigma|)
		(\frac{1}{2}+\frac{1}{8}m^2\sigma + \cdots) -
		\frac{1}{4}-\cdots  \right]  \nonumber \\
&  &  +{1\over {2 m^2}}\left[a_2 + \cdots \right] +
   {1\over {2 m^4}}\left[a_3 + \cdots \right] + \cdots \Biggr\}  .
\label{g1biscalar}
\end{eqnarray}
With the relations of Eq.(\ref{a0recursion}) and Eq.(\ref{anrecursion}) for the
$a_n$, the changes in $G^{(1)}(x,x')$ due to the presence of the
gauge field will be carried through the $a_n$. It is easily seen that
$G^{(1)}$ as expressed in Eq.(\ref{g1biscalar}) has at least a quadratic
divergence in the infinitesimal separation $\sigma^{,\mu}$ due to the presence
of the term proportional to $1/\sigma=2/(\sigma_{,\mu}\sigma^{,\mu})$.
Isolation of this and other divergences in $G^{(1)}(x,x')$ and its derivatives
are discussed in the next two sections.

\section{Coincidence limits}
\label{sec:coincidence}
In this section the coincidence limits $x'\rightarrow x$ of Eqs.(\ref{sigma}),
(\ref{VVM}), (\ref{a0recursion}), and (\ref{anrecursion}) and their
derivatives are developed. Synge's bracket notation \cite{Synge},
\begin{equation}
  [a(x,x')]\equiv \lim_{x'\to x}a(x,x'),
\label{Syngenotation}
\end{equation}
will be used to simplify the writing of the many limits needed.
The coincidence limits of $\sigma(x,x')$ and $\Delta^{1\over 2}(x,x')$
and their covariant derivatives are unaffected by the presence of the
gauge field and hence are unchanged from the results of Christensen. He finds
\cite{Christensenthesis},
\begin{eqnarray}
&&[\sigma^{;\mu}]=0,                  \label{sigmac1}     \\
&&[\sigma^{;\mu}{}_{\nu}]=g^{\mu}{}_{\nu}, \label{sigmac2} \\
&&[\sigma^{;\mu}{}_{\nu\sigma}]=0,         \label{sigmac3}  \\
&&[\sigma^{;\mu}{}_{\nu\rho\tau}]=S^{\mu}{}_{\nu\rho\tau}\equiv
 -{1\over 3}(R^{\mu}{}_{\rho\nu\tau}+R^{\mu}{}_{\tau\nu\rho}),
 \label{sigmac4} \\
&&[\sigma^{;\mu}{}_{\nu\alpha\tau\rho}]={3\over 4}
 (S^{\mu}{}_{\nu\alpha\tau;\rho}+S^{\mu}{}_{\nu\tau\rho;\alpha}+
  S^{\mu}{}_{\nu\rho\alpha;\tau}),
\label{sigmac5} \end{eqnarray}
and a six-derivative limit with 36 terms involving
$S^{\mu}{}_{\nu\rho\tau}$ which may be found in Ref.\cite{Christensenthesis}.
Note that a semicolon has been used here to emphasize that the covariant
derivatives contain only the Christoffel connections. Also,
\begin{eqnarray}
&&[\Delta^\frac{1}{2}]=1,   \label{deltac1}	\\
&&[\Delta^\frac{1}{2}{}_{;\mu}]=0, \label{deltac2}	\\
&&[\Delta^\frac{1}{2}{}_{;\mu\nu}]={1\over 6}R_{\mu\nu}, \label{deltac3}\\
&&[\Delta^\frac{1}{2}{}_{;\mu\nu\rho}]={1\over 12}
 (R_{\mu\nu;\rho}+R_{\nu\rho;\mu}+R_{\rho\mu;\nu}),
\label{deltac4} \end{eqnarray}
and a four-derivative limit with 12 terms which will not be shown
here.

Differentiating Eq.(\ref{a0recursion}) repeatedly we find,
\begin{eqnarray}
&&\sigma^{;\mu}a_{0|\mu}=0, \label{a0recurderiv1}  \\
&&\sigma^{;\mu\nu}a_{0|\mu}+ \sigma^{;\mu}a_{0|\mu}{}^{\nu}=0,
	\label{a0recurderiv2} \\
&&\sigma^{;\mu\nu\rho}a_{0|\mu}+\sigma^{;\mu\nu}a_{0|\mu}{}^{\rho}
 +\sigma^{;\mu\rho}a_{0|\mu}{}^{\nu}+\sigma^{;\mu}a_{0|\mu}{}^{\nu\rho}=0,
\label{a0recurderiv3} \end{eqnarray}
and so forth, where the slash is used to emphasize the gauge
covariant derivative being applied to $a_0$.
Taking the coincidence limit of Eqs.(35-37), and using Eqs. (26-30)
along with the commutation relation of the gauge covariant derivative,
\begin{equation}
  a_{n|\mu\nu}-a_{n|\nu\mu}=ieF_{\mu\nu}a_n.
\label{gaugecommutator} \end{equation}
we find (see also Ref.\cite{BV}),
\begin{eqnarray}
&&[a_0]=1,			\label{a0coin1}	\\
&&[a_{0|\mu}]=0,		\label{a0coin2}	\\
&&[a_{0|\mu\nu}]={1\over 2}ieF_{\mu\nu},	\label{a0coin3}	\\
&&[a_{0|\mu\nu\rho}]=
 {1\over 3}ie(F_{\mu\nu;\rho}+F_{\mu\rho;\nu}),	\label{a0coin4}
\end{eqnarray}
and
\begin{eqnarray}
&&[a_{0|\mu\nu\rho\tau}]={{ie}\over 4}
 [F_{\mu\nu;\rho\tau}+F_{\mu\rho;\nu\tau}+F_{\mu\tau;\nu\rho}] \nonumber\\
&&\qquad\qquad -{e^2\over 4}[F_{\mu\nu}F_{\rho\tau}+
	       F_{\mu\rho}F_{\nu\tau}+F_{\mu\tau}F_{\nu\rho}]	\nonumber\\
&&\qquad\qquad +\frac{ie}{4}[
   F_{\alpha\mu}   S^{\alpha}{}_{\nu\rho\tau}
  +F_{\alpha\nu}   S^{\alpha}{}_{\mu\tau\rho}
  +F_{\alpha\rho}S^{\alpha}{}_{\mu\tau\nu}
  +F_{\alpha\tau}  S^{\alpha}{}_{\mu\rho\nu} ].
\label{a0coin5} \end{eqnarray}
Doing the same for $a_1$ setting $n=0$ in Eq.(\ref{anrecursion}), yields,
\begin{eqnarray}
&&[a_1]=(\frac{1}{6}-\xi)R,       \label{a1coin1}             \\
&&[a_{1|\mu}]=\frac{1}{2}(\frac{1}{6}-\xi)R_{;\mu}+
  \frac{ie}{6}F_{\mu\alpha}{}^{;\alpha},  \label{a1coin2}
\end{eqnarray}
and
\begin{eqnarray}
 &&[a_{1|\mu\nu}]=
  (\frac{1}{20}-\frac{1}{3}\xi)R_{;\mu\nu}
  +\frac{1}{60}R_{\mu\nu;\alpha}{}^{\alpha}
  +\frac{1}{90}R^{\alpha\beta}R_{\alpha\nu\beta\mu}           \nonumber\\
&&\qquad\qquad
  -\frac{1}{45}R_{\mu\alpha}R^{\alpha}{}_{\nu}
  +\frac{1}{90}R_{\alpha\beta\gamma\mu}R^{\alpha\beta\gamma}{}_{\nu}
  -\frac{e^2}{6}F_{\mu\alpha}F_{\nu}{}^{\alpha}               \nonumber\\
&&\qquad\qquad -\frac{ie}{12}
  (F_{\mu\alpha}{}^{;\alpha}{}_{\nu}+F_{\nu\alpha}{}^{;\alpha}{}_{\mu})
  +\frac{ie}{2}(\frac{1}{6}-\xi)F_{\mu\nu}.
\label{a1coin3}
\end{eqnarray}
Finally, for $a_2$,
\begin{eqnarray}
[a_2]=&&-\frac{1}{180}R^{\alpha\beta}R_{\alpha\beta}
  +\frac{1}{180}R^{\alpha\beta\gamma\delta}R_{\alpha\beta\gamma\delta}
  +\frac{1}{6}(\frac{1}{5}-\xi)R_{;\alpha}{}^{\alpha}          \nonumber\\
&&+\frac{1}{2}(\frac{1}{6}-\xi)^2 R^2 -
   \frac{1}{12}F^{\alpha\beta}F_{\alpha\beta}.
\label{a2coincidence} \end{eqnarray}

With these coincidence limits, we will be able to construct the
quantities necessary to isolate the divergences in Eqs.(\ref{divcurrent},
\ref{divstress}).

\section{Covariant Expansions}
\label{sec:expansions}
Evaluating the divergences of $G^{(1)}$, the current, and the stress
tensor requires expanding all of the terms in $G^{(1)}$ and its
derivatives in power series about the point $x$ using the infinitesimal
separation vector $\sigma^{\mu} \equiv \sigma^{,\mu}$. The generic form
of the power series expansion of any bitensor $a^{\mu\nu\cdots}(x,x')$ is
\begin{equation}
  a^{\mu\nu\cdots}(x,x') = a0^{\mu\nu\cdots}+
  a1^{\mu\nu\cdots}{}_\alpha \sigma^{\alpha}+
  {1\over 2!}a2^{\mu\nu\cdots}{}_{\alpha\beta}
    \sigma^{\alpha}\sigma^{\beta}+\cdots.
\label{genericpower} \end{equation}
To evaluate the coefficients
$a0^{\mu\nu\cdots},a1^{\mu\nu\cdots}{}_\alpha,...$, we take the
coincidence limit $x'\rightarrow x$ of Eq.(\ref{genericpower}) and its
derivatives. For example, for the second rank bitensor
$a^{\mu\nu}(x,x')$, this yields,
\begin{eqnarray}
&&a0^{\mu\nu}=[a^{\mu\nu}],                 \label{powerc1}         \\
&&a1^{\mu\nu}{}_{\alpha}=[a^{\mu\nu}{}_{;\alpha}]
  -a0^{\mu\nu}{}_{;\alpha},                  \label{powerc2}         \\
&&a2^{\mu\nu}{}_{\alpha\beta}=[a^{\mu\nu}{}_{;\alpha\beta}]
  -2 a1^{\mu\nu}{}_{\alpha;\beta}-a0^{\mu\nu}{}_{;\alpha\beta},
\label{powerc3}\\
&&a3^{\mu\nu}{}_{\alpha\beta\gamma}=[a^{\mu\nu}{}_{;\alpha\beta\gamma}]\
  -3 a2^{\mu\nu}{}_{\alpha\beta;\gamma}-3 a1^{\mu\nu}{}_{\alpha;\beta\gamma}
  -a0^{\mu\nu}{}_{;\alpha\beta\gamma},
\label{powerc4} \end{eqnarray}
and so forth. The numeric factors on the right-hand side of
Eqs.(\ref{powerc1}-\ref{powerc4}) arise due to symmetrization on the dummy
indices $\alpha,\beta,...,$ in Eq.(\ref{genericpower}). Terms such as
$a1^{\mu\nu}{}_{\rho}S^{\rho}{}_{\alpha\beta\gamma}
\sigma^\alpha\sigma^\beta\sigma^\gamma$ do not contribute due to this
same symmetrization.

The expansions will have
bitensors constructed by taking primed derivatives of the biscalars in
$G^{(1)}$. These bitensors must be parallel transported back to the point
$x$ (see the form of Eq.(\ref{divcurrent}) and Eq.(\ref{divstress}))
in order to perform the expansion in Eq.(\ref{genericpower})
(See Ref.\cite{Christensen} for a more complete discussion of this point.)
For example,
\begin{equation}
  g^{\nu}{}_{\rho'}a^{;\mu\rho'}=
  a0^{\mu\nu}+a1^{\mu\nu}{}_\alpha \sigma^{\alpha}+\cdots \ .
\end{equation}
To evaluate the coincidence limits in Eqs.(\ref{powerc1}-\ref{powerc4})
for these primed derivatives, we will use Christensen's generalization
\cite{Christensen} of a theorem proved by Synge,
\begin{equation}
   [T_{\alpha_1\cdots\alpha_n\beta'_1\cdots\beta'_m;\mu'}]=
  -[T_{\alpha_1\cdots\alpha_n\beta'_1\cdots\beta'_m;\mu}]
  +[T_{\alpha_1\cdots\alpha_n\beta'_1\cdots\beta'_m}]_{;\mu},
\label{ChrSynge} \end{equation}
where $T_{\alpha_1\cdots\alpha_n\beta_1\cdots\beta_m}$ is any bitensor
with equal weight at both $x$ and $x'$ and whose coincidence limit and
derivative coincidence limits exist. We will also use the relation
\begin{equation}
 [g^{\mu}{}_{\nu';\alpha\beta\cdots}]\sigma^{\alpha}\sigma^{\beta}\cdots=0.
\label{bivectorderivs} \end{equation}
Applying these to $g^\mu{}_{\tau'}g^\nu{}_{\rho'}a^{|\tau'\rho'}$,
for example, yields,
\begin{eqnarray}
 a2^{\mu\nu}{}_{\alpha\beta} && =
 [(g^\mu{}_{\tau'}g^\nu{}_{\rho'}a^{|\tau'\rho'})_{|\alpha\beta}]
 -a0^{\mu\nu}{}_{;\alpha\beta}-a1^{\mu\nu}{}_{\alpha;\beta}
 -a1^{\mu\nu}{}_{\beta;\alpha},
\nonumber\\
&&=[a_{\alpha\beta}{}^{;\mu\nu}]-[a_{\alpha\beta}{}^{;\nu}]^{;\mu}
 -[a_{\alpha\beta}{}^{;\mu}]^{;\nu}+[a_{\alpha\beta}]^{;\mu\nu}
 -a0^{\mu\nu}{}_{;\alpha\beta}  \nonumber\\
&&-a1^{\mu\nu}{}_{\alpha;\beta}-a1^{\mu\nu}{}_{\beta;\alpha}
 +[g^\mu{}_{\sigma';\alpha\beta}]\hbox{\ terms,}
\label{a2umnlab}  \end{eqnarray}
where we have used the fact that primed derivatives commute with unprimed
derivatives, and terms containing objects such as $[g^\mu{}_{\sigma';\alpha}]$,
etc., have been grouped together since they contribute nothing to
the expansions due to (\ref{bivectorderivs}).

It can be seen from the expressions for $G^{(1)}$, the current, and the
stress tensor the order in $\sigma^\mu$ to which each biscalar expansion
must be carried. Applying this expansion method to the biscalars and
their gauge covariant derivatives with respect to both $x$ and $x'$, we
arrive at the following expansions,
\begin{eqnarray}
&&g^\mu{}_{\tau'}\sigma^{;\tau'}=-\sigma^{\mu},    \label{sp}\\
&&\sigma^{;\mu\nu}=g^{\mu\nu}
 -{1\over 3}R^\mu{}_\alpha{}^\nu{}_\beta \sigma^\alpha\sigma^\beta
 +{1\over 12}R^\mu{}_\alpha{}^\nu{}_{\beta;\gamma}
   \sigma^\alpha\sigma^\beta\sigma^\gamma                  \nonumber\\
&&\qquad -( {1\over 60}R^\mu{}_\alpha{}^\nu{}_{\beta;\gamma\delta}
   +{1\over 45}R^\rho{}_\alpha{}^\mu{}_\beta R_{\rho\gamma}{}^\nu{}_\delta)
  \sigma^\alpha\sigma^\beta\sigma^\gamma\sigma^\delta +\cdots,  \label{sdd}\\
&&g^\nu{}_{\rho'}\sigma^{;\mu\rho'}=-g^{\mu\nu}
 -{1\over 6}R^\mu{}_\alpha{}^\nu{}_\beta \sigma^\alpha\sigma^\beta
 +{1\over 12}R^\mu{}_\alpha{}^\nu{}_{\beta;\gamma}
 \sigma^\alpha\sigma^\beta\sigma^\gamma                     \nonumber\\
&&\qquad -( {1\over 40}R^\mu{}_\alpha{}^\nu{}_{\beta;\gamma\delta}
   +{7\over 360}R^\rho{}_\alpha{}^\mu{}_\beta R_{\rho\gamma}{}^\nu{}_\delta)
  \sigma^\alpha\sigma^\beta\sigma^\gamma\sigma^\delta +\cdots,  \label{sdp}\\
&&g^\mu{}_{\tau'}g^\nu{}_{\rho'}\sigma^{;\tau'\rho'}=g^{\mu\nu}
 -{1\over 3}R^\mu{}_\alpha{}^\nu{}_\beta \sigma^\alpha\sigma^\beta
 +{1\over 4}R^\mu{}_\alpha{}^\nu{}_{\beta;\gamma}
   \sigma^\alpha\sigma^\beta\sigma^\gamma                     \nonumber\\
&&\qquad -( {1\over 10}R^\mu{}_\alpha{}^\nu{}_{\beta;\gamma\delta}
   +{1\over 45}R^\rho{}_\alpha{}^\mu{}_\beta R_{\rho\gamma}{}^\nu{}_\delta)
  \sigma^\alpha\sigma^\beta\sigma^\gamma\sigma^\delta +\cdots,  \label{spp}\\
&&\Delta^{1\over 2}=1+{1\over 12}R_{\alpha\beta}\sigma^\alpha\sigma^\beta
 -{1\over 24}R_{\alpha\beta;\gamma}\sigma^\alpha\sigma^\beta\sigma^\gamma
  \nonumber\\
&&\qquad +({1\over 288}R_{\alpha\beta}R_{\gamma\delta}
 +{1\over 360}R^\rho{}_\alpha{}^\tau{}_\beta R_{\rho\gamma\tau\delta}
 +{1\over 80} R_{\alpha\beta;\gamma\delta})
  \sigma^\alpha\sigma^\beta\sigma^\gamma\sigma^\delta +\cdots,  \label{dp}\\
&&\Delta^{{1\over 2};\mu}={1\over 6}R^\mu{}_\alpha\sigma^\alpha
 -{1\over 24}(2R^\mu{}_{\alpha;\beta}-R_{\alpha\beta}{}^{;\mu})
  \sigma^\alpha\sigma^\beta
  +({1\over 40}R^\mu{}_{\alpha;\beta\gamma}               \nonumber\\
&&\qquad
  -{1\over 60}R_{\alpha\beta}{}^{;\mu}{}_\gamma
  +{1\over 90}R^{\rho\mu\tau}{}_\alpha R_{\rho\beta\tau\gamma}
  +{1\over 72}R^\mu{}_\alpha R_{\beta\gamma}
  +{1\over 360}R_\alpha{}^\rho R_{\rho\beta}{}^\mu{}_\gamma)
   \sigma^\alpha\sigma^\beta\sigma^\gamma +\cdots,          \label{dpd}\\
&&g^\mu{}_{\tau'}\Delta^{{1\over 2};\tau'}=
 -{1\over 6}R^\mu{}_\alpha\sigma^\alpha
 +{1\over 24}(2R^\mu{}_{\alpha;\beta}+R_{\alpha\beta}{}^{;\mu})
  \sigma^\alpha\sigma^\beta
 +(-{1\over 40}R^\mu{}_{\alpha;\beta\gamma}
  +{1\over 60}R_{\alpha\beta}{}^{;\mu}{}_\gamma                \nonumber\\
&&\qquad\quad
  -{1\over 90}R^{\rho\mu\tau}{}_\alpha R_{\rho\beta\tau\gamma}
  -{1\over 72}R^\mu{}_\alpha R_{\beta\gamma}
  +{11\over 360}R_\alpha{}^\rho R_{\rho\beta}{}^\mu{}_\gamma
  +{1\over 270}R^{\rho\tau\mu}{}_\alpha R_{\rho\beta\tau\gamma})
   \sigma^\alpha\sigma^\beta\sigma^\gamma +\cdots,             \label{dpp}\\
&&\Delta^{{1\over 2};\mu\nu}={1\over 6}R^{\mu\nu}
 +{1\over 12}(R^\mu{}_\alpha{}^{;\nu}+R^\nu{}_\alpha{}^{;\mu}
  -R^{\mu\nu}{}_{;\alpha})\sigma^\alpha                         \nonumber\\
&&\qquad
  +({1\over 40}R^{\mu\nu}{}_{;\alpha\beta}
  +{1\over 80}(R_{\alpha\beta}{}^{;\mu\nu}+R_{\alpha\beta}{}^{;\nu\mu})
  -{1\over 30}(R^\mu{}_{\alpha}{}^{;\nu}{}_\beta +
	       R^\nu{}_{\alpha}{}^{;\mu}{}_\beta )
  +{1\over 72}R_{\alpha\beta}R^{\mu\nu}
  +{1\over 36}R^\mu{}_\alpha R^\nu{}_\beta                       \nonumber\\
&&\qquad\quad
  +{1\over 360}(R^\mu{}_\rho R^\nu{}_\alpha{}^\rho{}_\beta +
		R^\nu{}_\rho R^\mu{}_\alpha{}^\rho{}_\beta )
  +{1\over 90}R^{\mu\rho\nu\tau}R_{\rho\alpha\tau\beta}
  +{1\over 180}(R^\rho{}_\alpha{}^{\tau\mu}R_\rho{}^\nu{}_{\tau\beta} +
		R^\rho{}_\alpha{}^{\tau\nu}R_\rho{}^\mu{}_{\tau\beta} )
    \nonumber\\
&&\qquad\quad
  +{1\over 180}(R^\rho{}_\alpha{}^{\tau\mu}R_\tau{}^\nu{}_{\rho\beta} +
		R^\rho{}_\alpha{}^{\tau\nu}R_\tau{}^\mu{}_{\rho\beta} )
  +{11\over 360}R_{\alpha\rho}
	       (R^{\rho\mu\nu}{}_\beta+R^{\rho\nu\mu}{}_\beta)  )
   \sigma^\alpha\sigma^\beta +\cdots,                     \label{dpdd}\\
&&g^\nu{}_{\rho'}\Delta^{{1\over 2};\mu\rho'}=
 -{1\over 6}R^{\mu\nu}
 +{1\over 12}(R^\mu{}_\alpha{}^{;\nu}-R^\nu{}_\alpha{}^{;\mu}
  -R^{\mu\nu}{}_{;\alpha})\sigma^\alpha                         \nonumber\\
&&\qquad
 +(-{1\over 40}R^{\mu\nu}{}_{;\alpha\beta}
  +{1\over 60}R_{\alpha\beta}{}^{;\mu\nu}
  +{1\over 30}(2 R^\mu{}_{\alpha;\beta}{}^\nu
   -5R^\mu{}_{\alpha}{}^{;\nu}{}_\beta+2R^\nu{}_{\alpha}{}^{;\mu}{}_\beta)
  -{1\over 72}R_{\alpha\beta}R^{\mu\nu}                           \nonumber\\
&&\qquad\quad
  -{1\over 36}R^\mu{}_\alpha R^\nu{}_\beta
  -{1\over 180}(R^\nu{}_\rho R^\mu{}_\alpha{}^\rho{}_\beta
		-11R^\mu{}_\rho R^\nu{}_\alpha{}^\rho{}_\beta)
  -{1\over 90}R^{\mu\rho\nu\tau}R_{\rho\alpha\tau\beta}         \nonumber\\
&&\qquad\quad
  -{1\over 90}R^\rho{}_\alpha{}^{\tau\mu}
   (R_\tau{}^\nu{}_{\rho\beta}+R_\rho{}^\nu{}_{\tau\beta})
  -{1\over 180}R_{\alpha\rho}
	 (R^\rho{}_\beta{}^{\mu\nu}+10R^{\rho\mu\nu}{}_\beta)  )
   \sigma^\alpha\sigma^\beta +\cdots,                          \label{dpdp}\\
&&g^\mu{}_{\tau'}g^\nu{}_{\rho'}\Delta^{{1\over 2};\tau'\rho'}=
 {1\over 6}R^{\mu\nu}
 -{1\over 12}(R^\mu{}_\alpha{}^{;\nu}+R^\nu{}_\alpha{}^{;\mu}
  +R^{\mu\nu}{}_{;\alpha})\sigma^\alpha                          \nonumber\\
&&\qquad
  +({1\over 40}R^{\mu\nu}{}_{;\alpha\beta}
  +{1\over 80}(R_{\alpha\beta}{}^{;\mu\nu}+R_{\alpha\beta}{}^{;\nu\mu})
  -{1\over 30}(R^\mu{}_{\alpha;\beta}{}^\nu+R^\nu{}_{\alpha;\beta}{}^\mu)
  +{1\over 12}(R^\mu{}_{\alpha}{}^{;\nu}{}_\beta +
	       R^\nu{}_{\alpha}{}^{;\mu}{}_\beta )             \nonumber\\
&&\qquad\quad
  +{1\over 72}R_{\alpha\beta}R^{\mu\nu}
  +{1\over 36}R^\mu{}_\alpha R^\nu{}_\beta
  -{11\over 360}(R^\mu{}_\rho R^\nu{}_\alpha{}^\rho{}_\beta +
		 R^\nu{}_\rho R^\mu{}_\alpha{}^\rho{}_\beta ) \nonumber\\
&&\qquad\quad
  +{1\over 180}R_{\rho\alpha\tau\beta}
	       (R^{\mu\rho\nu\tau}+R^{\nu\rho\mu\tau})
  +{1\over 180}(R^\rho{}_\alpha{}^{\tau\mu}R_\rho{}^\nu{}_{\tau\beta} +
		R^\rho{}_\alpha{}^{\tau\nu}R_\rho{}^\mu{}_{\tau\beta} )
    \nonumber\\
&&\qquad\quad
  +{1\over 180}(R^\rho{}_\alpha{}^{\tau\mu}R_\tau{}^\nu{}_{\rho\beta} +
		R^\rho{}_\alpha{}^{\tau\nu}R_\tau{}^\mu{}_{\rho\beta} )
  -{1\over 360}R_{\alpha\rho}
	       (R^{\rho\mu\nu}{}_\beta+R^{\rho\nu\mu}{}_\beta)  )
   \sigma^\alpha\sigma^\beta +\cdots,                         \label{dppp}\\
&&a_0=1,                                                      \label{a0}\\
&&a_0{}^{|\mu}=
 {ie\over 2}F^\mu{}_\alpha\sigma^\alpha
 -{ie\over 6}F^\mu{}_{\alpha;\beta}\sigma^\alpha\sigma^\beta
 +{ie\over 24}(F^\mu{}_{\alpha;\beta\gamma} +
	       F_{\rho\alpha}R^\mu{}_\beta{}^\rho{}_\gamma)
   \sigma^\alpha\sigma^\beta\sigma^\gamma +\cdots,              \label{a0d}\\
&&g^\mu{}_{\tau'}a_0{}^{|\tau'}=
 {ie\over 2}F^\mu{}_\alpha\sigma^\alpha
 -{ie\over 3}F^\mu{}_{\alpha;\beta}\sigma^\alpha\sigma^\beta
 +{ie\over 24}(3F^\mu{}_{\alpha;\beta\gamma} +
		F_{\rho\alpha}R^\mu{}_\beta{}^\rho{}_\gamma)
   \sigma^\alpha\sigma^\beta\sigma^\gamma +\cdots,              \label{a0p}\\
&&a_0{}^{|\mu\nu}={ie\over 2}F^{\mu\nu}
 +{ie\over 6}(F^\mu{}_\alpha{}^{;\nu}+F^\nu{}_\alpha{}^{;\mu})\sigma^\alpha
 +({ie\over 24}(F_\alpha{}^\mu{}_{;\beta}{}^\nu +
		F_\alpha{}^\nu{}_{;\beta}{}^\mu )
   -{{e^2}\over 4}F_\alpha{}^\mu F_\beta{}^\nu                  \nonumber\\
&&\qquad\quad
   +{ie\over 12}(F_\rho{}^\mu R^\nu{}_\alpha{}^\rho{}_\beta +
				 F_\rho{}^\nu R^\mu{}_\alpha{}^\rho{}_\beta )
   -{ie\over 24}F_{\rho\alpha}
     (R^{\mu\rho\nu}{}_\beta+R^{\nu\rho\mu}{}_\beta) )
    \sigma^\alpha\sigma^\beta +\cdots,                        \label{a0dd}\\
&&g^\nu{}_{\rho'}a_0{}^{|\mu\rho'}=-{ie\over 2}F^{\mu\nu}
 +{ie\over 6}(F^{\mu\nu}{}_{;\alpha}+F^\mu{}_\alpha{}^{;\nu})\sigma^\alpha
 +(-{ie\over 24}F^{\mu\nu}{}_{;\alpha\beta}
   +{ie\over 12}F_\alpha{}^\mu{}_{;\beta}{}^\nu
   -{{e^2}\over 4}F_\alpha{}^\mu F_\beta{}^\nu                \nonumber\\
&&\qquad\quad
   +{ie\over 24}(3F_\rho{}^\mu R^\nu{}_\alpha{}^\rho{}_\beta -
				  F_\rho{}^\nu R^\mu{}_\alpha{}^\rho{}_\beta )
   +{ie\over 12}F_{\rho\alpha}R^{\nu\rho\mu}{}_\beta  )
    \sigma^\alpha\sigma^\beta +\cdots,                        \label{a0dp}\\
&&g^\mu{}_{\tau'}g^\nu{}_{\rho'}a_0{}^{|\tau'\rho'}=
 {ie\over 2}F^{\mu\nu}
 +{ie\over 6}(F^\mu{}_\alpha{}^{;\nu}+F^\nu{}_\alpha{}^{;\mu})\sigma^\alpha
 +({3ie\over 8}(F_\alpha{}^\mu{}_{;\beta}{}^\nu +
		F_\alpha{}^\nu{}_{;\beta}{}^\mu )
   -{{e^2}\over 4}F_\alpha{}^\mu F_\beta{}^\nu               \nonumber\\
&&\qquad\quad
   +{ie\over 4}(F_\rho{}^\mu R^\nu{}_\alpha{}^\rho{}_\beta +
				 F_\rho{}^\nu R^\mu{}_\alpha{}^\rho{}_\beta )
   +{3ie\over 8}F_{\rho\alpha}
     (R^{\mu\rho\nu}{}_\beta+R^{\nu\rho\mu}{}_\beta) )
    \sigma^\alpha\sigma^\beta +\cdots,                     \label{a0pp}\\
&&a_1=({1\over 6}-\xi)R
 +(-{1\over 2}({1\over 6}-\xi)R_{;\alpha}+{ie\over 6}F^\rho{}_{\alpha;\rho})
   \sigma^\alpha
 +(-{1\over 90}R_{\alpha\rho}R^\rho{}_\beta
   +{1\over 180}R^{\rho\tau}R_{\rho\alpha\tau\beta}
   +{1\over 120}R_{\alpha\beta;\rho}{}^\rho                 \nonumber\\
&&\qquad\quad
   +{1\over 180}R_{\rho\tau\kappa\alpha}R^{\rho\tau\kappa}{}_\beta
   +({1\over 40}-{1\over 6}\xi)R_{;\alpha\beta}
   -{ie\over 12}F^\rho{}_{\alpha;\rho\beta}
   -{{e^2}\over 12}F^\rho{}_\alpha F_{\rho\beta} )
    \sigma^\alpha\sigma^\beta +\cdots,                      \label{a1}\\
&&a_1{}^{|\mu}={1\over 2}({1\over 6}-\xi)R^{;\mu}
 +{ie\over 6}F^{\rho\mu}{}_{;\rho}
 +(-{1\over 45}R^\mu{}_\rho R^\rho{}_\alpha
   +{1\over 90}R^{\rho\tau}R^\mu{}_{\rho\alpha\tau}
   +{1\over 90}R_{\rho\tau\kappa}{}^\mu R^{\rho\tau\kappa}{}_\alpha
     \nonumber\\
&&\qquad\quad
   +{1\over 60}R^\mu{}_{\alpha;\rho}{}^\rho
   +{1\over 6}(\xi-{1\over 5})R^{;\mu}{}_\alpha
   +{ie\over 12}(F_{\rho\alpha}{}^{;\rho\mu}-F^{\rho\mu}{}_{;\rho\alpha})
   +{ie\over 2}({1\over 6}-\xi)RF^\mu{}_\alpha )
     \sigma^\alpha+\cdots,                            \label{a1d}\\
&&g^\mu{}_{\tau'}a_1{}^{|\tau'}={1\over 2}({1\over 6}-\xi)R^{;\mu}
 -{ie\over 6}F^{\rho\mu}{}_{;\rho}  \nonumber\\
&&\qquad
 +( {1\over 45}R^\mu{}_\rho R^\rho{}_\alpha
   -{1\over 90}R^{\rho\tau}R^\mu{}_{\rho\alpha\tau}
   -{1\over 90}R_{\rho\tau\kappa}{}^\mu R^{\rho\tau\kappa}{}_\alpha
   -{1\over 60}R^\mu{}_{\alpha;\rho}{}^\rho  \nonumber\\
&&\qquad\quad
   +{1\over 3}(\xi-{3\over 20})R^{;\mu}{}_\alpha
   +{ie\over 12}(F_{\rho\alpha}{}^{;\rho\mu}-F^{\rho\mu}{}_{;\rho\alpha})
   +{ie\over 2}({1\over 6}-\xi)RF^\mu{}_\alpha )
     \sigma^\alpha+\cdots,                                 \label{a1p}\\
&&a_1{}^{|\mu\nu}=
 -{1\over 45}R^\mu{}_\rho R^{\nu\rho}
 +{1\over 90}R^{\rho\tau}R^\mu{}_\rho{}^\nu{}_\tau
 +{1\over 90}R_{\rho\tau\kappa}{}^\mu R^{\rho\tau\kappa\nu}
 +{1\over 60}R^{\mu\nu}{}_{;\rho}{}^\rho
 +({1\over 20}-{1\over 3}\xi)R^{;\mu\nu}                 \nonumber\\
&&\qquad\quad
 +{ie\over 12}(F_\rho{}^{\mu;\rho\nu}+F_\rho{}^{\nu;\rho\mu})
 +{ie\over 2}({1\over 6}-\xi)RF^{\mu\nu}
 -{{e^2}\over 6}F_\rho{}^\mu F^{\rho\nu} +\cdots,       \label{a1dd}\\
&&g^\nu{}_{\rho'}a_1{}^{|\mu\rho'}=
  {1\over 45}R^\mu{}_\rho R^{\nu\rho}
 -{1\over 90}R^{\rho\tau}R^\mu{}_\rho{}^\nu{}_\tau
 -{1\over 90}R_{\rho\tau\kappa}{}^\mu R^{\rho\tau\kappa\nu}
 -{1\over 60}R^{\mu\nu}{}_{;\rho}{}^\rho
 +{1\over 6}({1\over 5}-\xi)R^{;\mu\nu}                  \nonumber\\
&&\qquad\quad
 +{ie\over 12}(F_\rho{}^{\mu;\rho\nu}-F_\rho{}^{\nu;\rho\mu})
 -{ie\over 2}({1\over 6}-\xi)RF^{\mu\nu}
 +{{e^2}\over 6}F_\rho{}^\mu F^{\rho\nu} +\cdots,        \label{a1dp}\\
&&g^\mu{}_{\tau'}g^\nu{}_{\rho'}a_1{}^{|\tau'\rho'}=
 -{1\over 45}R^\mu{}_\rho R^{\nu\rho}
 +{1\over 90}R^{\rho\tau}R^\mu{}_\rho{}^\nu{}_\tau
 +{1\over 90}R_{\rho\tau\kappa}{}^\mu R^{\rho\tau\kappa\nu}
 +{1\over 60}R^{\mu\nu}{}_{;\rho}{}^\rho
 +({1\over 20}-{1\over 3}\xi)R^{;\mu\nu}   \nonumber\\
&&\qquad\quad
 -{ie\over 12}(F_\rho{}^{\mu;\rho\nu}+F_\rho{}^{\nu;\rho\mu})
 +{ie\over 2}({1\over 6}-\xi)RF^{\mu\nu}
 -{{e^2}\over 6}F_\rho{}^\mu F^{\rho\nu} +\cdots, \label{a1pp}\\
&&\hbox{\ and}  \nonumber\\
&&a_2=
 {1\over 180}(R^{\rho\tau\kappa\lambda}R_{\rho\tau\kappa\lambda}
			 -R^{\rho\tau}R_{\rho\tau})
 +{1\over 6}({1\over 5}-\xi)R_{;\rho}{}^\rho
 +{1\over 2}({1\over 6}-\xi)^2 R^2
 -{{e^2}\over 12}F_{\rho\tau}F^{\rho\tau} +\cdots.  \label{a2}
\end{eqnarray}
A semicolon has been used for all of the derivatives on the right hand
sides of Eqs.(\ref{sp}-\ref{a2}). All of the affected terms are gauge
invariant and thus do not need the connection $A^\mu$ in their
derivatives. The expansions with either two unprimed or two primed
derivatives have been symmetrized according to,
\begin{equation}
 a^{|\mu\nu}={1\over 2}(a^{|\mu\nu}+a^{|\nu\mu}+\chi^{\mu\nu}a),
\end{equation}
where $\chi^{\mu\nu}=ieF^{\mu\nu}$ for the $a_n$ and $\chi^{\mu\nu}=0$
for the other biscalars. Note that quantities such as
$g^\nu{}_{\rho'}(\sigma^{-1})^{;\mu\rho'}$ may only have their numerators
expanded using Eqs.(\ref{sp}-\ref{a2}) since they diverge in the
coincidence limit necessary for applying the expansion iterations in
Eqs.(\ref{powerc1}-\ref{powerc4}):
\begin{eqnarray}
 g^\nu{}_{\rho'}(\sigma^{-1})^{;\mu\rho'} &=&
 g^\nu{}_{\rho'}
   (2\sigma^{-3}\sigma^{;\mu}\sigma^{;\rho'}-\sigma^{-2}\sigma^{;\mu\rho'})
   \nonumber\\
 &=&
 -2\sigma^{-3}\sigma^{;\mu}\sigma^{;\nu}-\sigma^{-2}(-g^{\mu\nu}-\cdots).
\end{eqnarray}

\section{Results}
\label{sec:results}
Substituting the expansions in Eqs.(\ref{sp}-\ref{a2}) for the biscalars
into Eq.(\ref{g1biscalar}) and collecting terms of like powers of
$\sigma^\mu$ yields,
\begin{eqnarray}
 4\pi^2 G^{(1)}(x,x')=&&{2\over(\sigma^\rho\sigma_\rho)}
 +[m^2-({1\over 6}-\xi)R]
  [\gamma+{1\over 2}ln|{1\over 4}m^2(\sigma^\rho\sigma_\rho)|]
 -{1\over 2}m^2     \nonumber\\
&&+{1\over 6}R_{\alpha\beta}
  {{\sigma^\alpha\sigma^\beta}\over(\sigma^\rho\sigma_\rho)}
 +{1\over {2m^2}}[-{1\over 180}R^{\rho\tau}R_{\rho\tau}
  +{1\over 180}R^{\rho\tau\kappa\lambda}R_{\rho\tau\kappa\lambda}
  +{1\over 6}({1\over 5}-\xi)R_{;\rho}{}^\rho \nonumber\\
&&+{1\over 2}({1\over 6}-\xi)^2 R^2]+\vartheta\left({1\over{m^4}}\right).
\label{g1answer} \end{eqnarray}
This is the same as the result derived by Christensen for a real scalar
field.

We differentiate Eq.(\ref{g1biscalar}) to find $G^{(1)|\mu}$ and
$G^{(1)|\mu'}$, using the bivector of parallel transport
$g^\mu{}_\sigma'$ to construct the expectation value of the current defined
in Eq.(\ref{divcurrent}). Substituting
the appropriate expansions from Eqs.(\ref{sp}-\ref{a2}) and collecting
like powers of $\sigma^\mu$ isolates the terms in Eq.(\ref{divcurrent}) which
will diverge linearly and those which will remain finite as the points are
brought together:
\begin{equation}
  \langle\underline{j}^\mu(x,x')\rangle_{linear}={1\over{4\pi^2}}
 {{e^2 \sigma^\alpha F_\alpha{}^\mu}\over(\sigma^\rho\sigma_\rho)},
\label{linearcurrent} \end{equation}
and,
\begin{equation}
  \langle\underline{j}^\mu(x,x')\rangle_{finite}={1\over{4\pi^2}}
 {{e^2 \sigma^\alpha\sigma^\beta F^\mu{}_{\alpha;\beta}}
 \over(2\sigma^\rho\sigma_\rho)}+\vartheta\left({1\over{m^2}}\right).
\label{finitecurrent} \end{equation}

Finally, we form $G^{(1)|\mu\nu},G^{(1)|\mu'\nu},G^{(1)|\mu\nu'}$, and
$G^{(1)|\mu'\nu'}$ by differentiating Eq.(\ref{g1biscalar}). These
are then used to form the components of the stress energy tensor of
Eq.(\ref{divstress}). Substituting the expansions from
Eqs.(\ref{sp}-\ref{a2}), and collecting like powers of $\sigma^\mu$,
yields the following expressions, with the points split,
\begin{eqnarray}
\langle\underline{T}^{\mu\nu}\rangle_{quartic}&=&{1\over{2\pi^2}}
 {1\over(\sigma^\rho\sigma_\rho)^2}
 \left[g^{\mu\nu}
  -4{{\sigma^\mu\sigma^\nu}\over(\sigma^\rho\sigma_\rho)}
 \right], \label{quardiv}                                       \\
\langle\underline{T}^{\mu\nu}\rangle_{quadratic}&=&{1\over{4\pi^2}}
 {1\over(\sigma^\rho\sigma_\rho)} \times    \nonumber\\
&&  \Bigg[{1\over 3}(R^\mu{}_\alpha\sigma^\nu+R^\nu{}_\alpha\sigma^\mu)
   {{\sigma^\alpha}\over(\sigma^\rho\sigma_\rho)}
   -{2\over 3}R_{\alpha\beta}
    {{\sigma^\alpha\sigma^\beta\sigma^\mu\sigma^\nu}
    \over(\sigma^\rho\sigma_\rho)^2}  \nonumber\\
&&-{1\over 2}m^2\left[g^{\mu\nu}-2{{\sigma^\mu\sigma^\nu}
    \over(\sigma^\rho\sigma_\rho)}\right]
  -({1\over 6}-\xi) \Bigg\{
   R^{\mu\nu}-{1\over 2}R\left[g^{\mu\nu}-2{{\sigma^\mu\sigma^\nu}
    \over(\sigma^\rho\sigma_\rho)}\right]  \nonumber\\
&&+2(g^{\mu\nu}R_{\alpha\beta}-R^\mu{}_\alpha{}^\nu{}_\beta)
   {{\sigma^\alpha\sigma^\beta}\over(\sigma^\rho\sigma_\rho)}
   \Bigg\}
  \Bigg],
\label{quaddiv} \\
\langle\underline{T}^{\mu\nu}\rangle_{linear}&=&{1\over{4\pi^2}}
 \times                                                    \nonumber\\
&&\Bigg[{1\over 12}(R^{\mu\nu}-{1\over 4}Rg^{\mu\nu})_{;\alpha}
   {{\sigma^\alpha}\over(\sigma^\rho\sigma_\rho)}
  -{1\over 6}(R^\mu{}_\alpha g^\nu{}_\beta+R^\nu{}_\alpha g^\mu{}_\beta
   -{1\over 4}R_{\alpha\beta}g^{\mu\nu})_{;\gamma}
   {{\sigma^\alpha\sigma^\beta\sigma^\gamma}\over
   (\sigma^\rho\sigma_\rho)^2}  \nonumber\\
&& -({1\over 6}-\xi)\left\{
  (R^\mu{}_\alpha{}^\nu{}_\beta-R_{\alpha\beta}g^{\mu\nu})_{;\gamma}
   {{\sigma^\alpha\sigma^\beta\sigma^\gamma}\over
   (\sigma^\rho\sigma_\rho)^2}
  +{1\over 4}R_{;\alpha}{{\sigma^\alpha}\over(\sigma^\rho\sigma_\rho)}
   \left[g^{\mu\nu}-2{{\sigma^\mu\sigma^\nu}\over(\sigma^\rho\sigma_\rho)}
   \right]\right\}  \nonumber\\
&& -{1\over 12}R_{\alpha\beta;\gamma}
  {{\sigma^\alpha\sigma^\beta\sigma^\gamma}\over
   (\sigma^\rho\sigma_\rho)^2}
  \left[g^{\mu\nu}-4{{\sigma^\mu\sigma^\nu}\over
    (\sigma^\rho\sigma_\rho)}\right]
  +({1\over 6}-\xi)^2\left[{3\over 4}R_{;\alpha}{{\sigma^\alpha}\over
   (\sigma^\rho\sigma_\rho)}g^{\mu\nu}\right]
   \Bigg],
\label{lindiv}\\
\langle\underline{T}^{\mu\nu}\rangle_{logarithmic}&=&{1\over{4\pi^2}}
 [\gamma+{1\over 2}ln|{1\over 4}m^2(\sigma^\rho\sigma_\rho)|]\times
\nonumber\\
&&\Bigg[ {1\over 60}(R^{\rho\mu\tau\nu}R_{\rho\tau}-
   {1\over 4}R^{\rho\tau}R_{\rho\tau}g^{\mu\nu})
  -{1\over 2}({1\over 6}-\xi)
   \left[m^2(R^{\mu\nu}-{1\over 2}Rg^{\mu\nu})\right]	\nonumber\\
&&+{1\over 120}R^{\mu\nu}{}_{;\rho}{}^\rho
  -{1\over 180}R(R^{\mu\nu}-{1\over 4}Rg^{\mu\nu})
  -{1\over 360}R^{;\mu\nu}-{1\over 720}R_{;\rho}{}^\rho g^{\mu\nu} \nonumber\\
&&-{1\over 2}({1\over 6}-\xi)^2\left[
    R^{;\mu\nu}-R(R^{\mu\nu}-{1\over 4}Rg^{\mu\nu})
    -R_{;\rho}{}^\rho g^{\mu\nu} \right]  \nonumber\\
&&-{1\over 8}m^4 g^{\mu\nu}
  -{e^2\over 12}F^{\rho\mu}F_\rho{}^\nu
  +{e^2\over 48}g^{\mu\nu} F^{\rho\tau}F_{\rho\tau} \Bigg],
\label{logdiv}
\end{eqnarray}
and
\begin{eqnarray}
\langle\underline{T}^{\mu\nu}\rangle_{finite}&=&
\langle\underline{T}^{\mu\nu}\rangle_{finite,Christensen} + \nonumber\\
&&{e^2\over{4\pi^2}}\Bigg[{{\sigma^\mu\sigma^\nu}\over
	{12(\sigma^\rho\sigma_\rho)^2}}
   \sigma^\alpha\sigma^\beta F_\alpha{}^\gamma F_{\beta\gamma}
  +{1\over{12(\sigma^\rho\sigma_\rho)}}
   (\sigma^\alpha\sigma^\mu F_\alpha{}^\beta F_\beta{}^\nu +
    \sigma^\alpha\sigma^\nu F_\alpha{}^\beta F_\beta{}^\mu )   \nonumber \\
&&+{1\over(\sigma^\rho\sigma_\rho)}(\xi-{1\over 4})
   \sigma^\alpha\sigma^\beta F_\alpha{}^\mu F_\beta{}^\nu
  +{1\over(\sigma^\rho\sigma_\rho)}(\xi-{5\over 24})g^{\mu\nu}
   \sigma^\alpha\sigma^\beta F_\alpha{}^\gamma F_{\beta\gamma} \nonumber \\
&&-{1\over 96}\left( g^{\mu\nu}-2{{\sigma^\mu\sigma^\nu}\over
   (\sigma^\rho\sigma_\rho)} \right) F^{\alpha\beta}F_{\alpha\beta}
  +\vartheta\left({1\over{m^2}}\right) \Bigg].  \label{finite}
\end{eqnarray}
For brevity, only the terms due to the electromagnetic field are included
in Eq.(\ref{finite}). The terms
$\langle\underline{T}^{\mu\nu}\rangle_{finite,Christensen}$ may be found
in Ref.\cite{Christensen}.

\section{Discussion}
\label{sec:discussion}

The complex scalar field is constructed from two real fields according to
$\phi(x)={1\over\sqrt{2}}\left(\phi_{1}(x)+i\phi_{2}(x)\right)$. From the
form of the Hadamard function Eq.(\ref{G1}), it is evident that $G^{(1)}$
for the complex scalar field will be the sum of $G^{(1)}$ for two
real scalar fields. Thus, the agreement of Eq.(\ref{g1answer}) with
Christensen's original result is to be expected, with both containing the
same quadratic and logarithmic divergences, with their corresponding
direction-dependent ($\sigma^\mu$-dependent) and direction-independent terms.

The divergence of Eqs.(\ref{linearcurrent}) and (\ref{finitecurrent})
cannot be taken directly in order to verify the point-splitting procedure
does not violate the conservation of current. This would
involve taking the divergence with respect to $x^\alpha$
of the separation vector $\sigma^{\mu}(x,x')$, which depends on both
$x^\alpha$ and $x'^\alpha$.
The correct procedure for verifying conservation of current requires
evaluating the divergence of the classical current within the
point-splitting regime. The divergence of the classical current,
\begin{equation}
 j^\mu{}_{;\mu}=ie[(D^\mu \phi)\phi^*-\phi(D^\mu \phi)^*]_{;\mu},
\label{classdivergence1}\end{equation}
may be written as,
\begin{equation}
 j^\mu{}_{;\mu}={ie\over 2}
  \left[ \{D_\mu D^\mu\phi,\phi^* \}-\{D_\mu D^\mu\phi,\phi^* \}^*
  \right]
 ={ie\over 2}
  \left[ \{\phi^{|\mu}{}_\mu,\phi^* \}-\{\phi^{|\mu}{}_\mu,\phi^* \}^*
  \right].
\label{classdivergence2}\end{equation}
Making the transition from classical to quantum fields and applying the
point-splitting procedure to the right-hand-side of Eq.(\ref{classdivergence2})
yields the proper expression to examine, namely,
\begin{equation}
\langle\underline{j}^\mu{}_{;\mu}(x) \rangle =
  \lim_{x'\to x}{ie\over 4}\left[
  \left( G^{(1)|\mu}{}_\mu+
    g^{\mu}{}_{\tau'}g_{\mu\rho'}G^{(1)|\tau'\rho'} \right) -
  \left( G^{(1)|\mu}{}_\mu+
    g^{\mu}{}_{\tau'}g_{\mu\rho'}G^{(1)|\tau'\rho'} \right)^*\right].
\label{currentdivergence} \end{equation}
This may be seen to vanish to all orders by using the expansions
in Eqs.(\ref{sp}-\ref{a2}), ensuring that the current is a conserved
quantity.

The linear divergence of Eq.(\ref{linearcurrent}) is to be expected by a
straightforward analysis of the divergences present in scalar
electrodynamics \cite{Bjorken,Schweber}. The interaction terms in the
Lagrangian for the charged scalar field, $L_{I}\equiv j^\mu A_\mu$, may
be written in the form,
\begin{equation}
 L_{I}={ie\over 2}\left[
   \{\partial^\mu\phi,\phi^*\}-\{\partial^\mu\phi,\phi^*\}^* \right]A_\mu
   -e^2 A^\mu A_\mu\{\phi,\phi^*\}.
\label{scalarinteract} \end{equation}
The first term gives rise to a 3-point Feynman graph, while the second
term has a 4-point graph. The degree of divergence $D$ present in scalar
electrodynamics is given by,
\begin{equation}  D=4-P_e - Q_e,
\label{scalardiv} \end{equation}
where $P_e$ and $Q_e$ are the number of external scalar and photon lines,
respectively. The values $D=0,1,2,...$, imply logarithmic, linear,
quadratic,..., divergences, while $D<0$ implies the interaction is
finite. DeWitt has pointed out for the generalized Yang-Mills field in
the presence of the gravitational field that the simplest possible
counterterm will always be the most divergent \cite{DeWitt}.
The simpler term represented by the 3-point graph indicates a degree of
divergence $D=1$. This linear divergence is in contrast to the well-known
logarithmic divergence of the fermion field current \cite{Schwinger}. The
fermion field has an interaction Lagrangian given by,
\begin{equation}
 L_{I}={e\over 2}[\overline{\psi},\gamma^\mu\psi]A_\mu,
\label{fermioninteract} \end{equation}
where the sum over the indices of the spinors and the gamma matrices is
understood. The degree of divergence for spinor electrodynamics is given
by,
\begin{equation}  D=4-{3\over 2}F_e - Q_e,
\label{spinordiv} \end{equation}
where $F_e$ and $Q_e$ are the number of external fermion and photon lines,
respectively. The single 3-point graph indicated by
(\ref{fermioninteract}) would thus have $D=0$, an expected logarithmic
divergence.

The current and stress energy tensor are independent of the sign of the
charge carrier with both having electromagnetic terms proportional to
$e^2$. One factor of $e$ in the current is explicit in
Eq.(\ref{divcurrent}). The other factor comes from the derivation of
the expansions of $G(1)^{|\mu}$, etc. This originates with the gauge
commutation relation, Eq.(\ref{gaugecommutator}), and is seen to carry
one factor of $eF$ through to the first derivative expansions of the
$a_n$ in Eqs.(\ref{sp}-\ref{a2}). The current is a third order
quantity with units of $(length)^{-3}$.
By counting powers, the linearly diverging term would thus
require one factor of the second order $eF$, while the finite term would
require the third order derivative of $eF$. Eqs. (\ref{linearcurrent})
and (\ref{finitecurrent}) are thus in the only possible form and, aside
from numerical factors, could have been deduced immediately from simple
power counting.

The structure of the electromagnetic counterterms for the stress energy
tensor may be deduced in like manner. They must be independent of the
sign of the charge and thus proportional to $e^2$. There are no explicit
factors of $e$ in Eq.(\ref{divstress}), so they must arise in the
derivations of the expansions of $G(1)^{|\mu\nu}$, etc., by virtue of the
gauge commutator. The simplest term expected is the fourth order
$(eF)^2$. The fourth order stress energy tensor would only allow such a
term to be present in the logarithmic counterterm. This is in agreement with
DeWitt's result for the generalized Yang-Mills field \cite{DeWitt}. The
two possible fourth order combinations of the  gravitational and
electromagnetic field tensors are $g^{\mu\nu} e^2
F^{\rho\tau}F_{\rho\tau}$ and $e^2 F^{\rho\mu}F_{\rho}{}^\nu$. The
point-splitting procedure yields the numerical factors for these terms.

It is to be expected that some infinite counterterms involving only the
electromagnetic field would remain even in the flat space limit. The
present work yields some of those terms. It is not surprising that there
are no cross terms here involving the electromagnetic field and
curvature. The lowest order of such cross terms would be sixth order. The
point-splitting procedure is an expansion in inverse powers of $m$
\cite{DeWitt} and the current and stress energy tensor in the present
case are truncated at order $m^0$. Sixth order cross terms such as
$(eF)^2 R$ are expected when the expansion is carried out to
order $m^{-2}$ \cite{Hunp}.

\acknowledgements
R.\ H.\ would like to thank Emil Mottola for helpful discussions. The work
of R.\ H.\ was supported by the Department of Energy under Cooperative
Agreement \# DE-FC02-91ER75681, Amnd. \# 002; the work of W.\ A.\ H.\
was supported in part by the National Science Foundation, Grant \# PHY92-07903.

\end{document}